\documentclass[12pt]{article}

\title{Late periods of the condensation process }
\author{Victor Kurasov}
\date{Department of Computational Physics,
St. Petersburg State University,
St, Petersburg, Russia\footnote{Victor\_Kurasov@yahoo.com}}
\begin{document}

\maketitle
\begin{abstract}
The full evolution during the late periods of the
condensation process is described in the analytical form.
The process is split into several periods and for every
period the simple approximate solution is given.
\end{abstract}

\section{Introduction}

The first order phase transition is characterized by
the temporal duration - the process of the condensation lasts
in time and consists of several characteristic periods.
The kinetics of the first order
phase transition is one of the actual problems in the
phase transformations.

Ordinary it is supposed that the final stage (period)
of the phase transition is the stage of coalescence.
This name goes from the theory given by Lifshic and
Slezov \cite{slez}.
This terminology is not absolutely correct - formally
the coalescence means the adhesion of the embryos. But
the Lifshic-Slezov (LS) consideration does not take
into account the adhesion, the evolution in the LS
picture is the competition between the embryos through
the exhaustion of the vapor environment. One has to
clarify this difference. In the case of adhesion we
shall speak about the coagulation and when the
Lifshic-Slezov mechanism takes place we shall speak
about the over-condensation.

The over-condensation means the competition between the
already formed embryos when the embryos of the relatively
big sizes
eat the embryos of the relatively small sizes
not directly but through
the
exhaustion of the metastable old phase until such a
degree that the mentioned small embryos become
the pre-critical
ones and begin to eject the molecules in the vapor.
These embryos begin, thus, to disappear.

Beside this opportunity it is possible to see in the
open systems the situation when all embryos continue to
grow until all volume of the system will be transformed
into a new phase. Certainly this situation takes place
when there is an effective source of the substance in
the old phase. This is also an alternative possibility
for the final of the phase transition.
So, we see that the final stages of the
phase transformation can be very different.

Here we shall speak about the over-condensation and
consider the closed system. The effect of coagulation
is not taken here into account. The temperature of the
system is supposed to be fixed. The mentioned
restrictions are not the crucial ones.

The methodology of
analysis of such systems is given by Lifshic and Slezov
in \cite{slez}.
Strictly speaking the results of Lifshic and Slezov can
not be directly applied for the systems with the
diffusion regime of the substance exchange
(as it was done in \cite{slez}) because in
this case around every embryo
there will be  a profile of the metastable
(old) phase substance. The profiles
around different  embryos overlap. This changes
the rates of growth. Then one has
to consider an interaction between the density
profiles. So, one can not directly
take the integral form of the
balance equation as it is done in the theory of
Lifshic-Slezov. The last task to describe the
interaction of profiles is extremely difficult and the
precise analytical solution of this problem is hardly
possible.

The ignorance of the diffusion profiles leads to the
necessity to consider the situation with the free
molecular regime of interaction when there is no
profiles. This was done by Wagner \cite{wagner} and we
shall follow this regime of the substance exchange
below.

The fundamental fact is that the LS asymptotics was
confirmed experimentally (see \cite{exp1}, \cite{exp2},
\cite{kukbook}). Here one has to stress that according
to the notation made in \cite{kuk} the accuracy is not
very high - one can speak only about the qualitative
confirmation of the form of the spectrum and about the
precise confirmation of the power-like law of evolution
of parameters of the spectrum (in some sense the
power-like law is rather evident and it can be derived
from some simple qualitative remarks). Also it is
necessary to stress that according to the references
from \cite{kuk} the form of the size spectrum is more
sharp than it follows from the leading term of the LS
theory.

The next step in modification of the LS approach was
done by Osipov and Kukushkin in \cite{kuk} where the
"regular asymptotic" was constructed.
Since the first two terms in asymptotic decomposition
in the LS
theory are universal (independent from the initial size
spectrum) it is possible to introduce such variables
(coordinates) where
already the initial approximation is equivalent to
the first
 two terms of the LS decomposition.
So, the modification of Osipov and Kukushkin seems to
be natural. But this modification
is important because it allows to
speak on the level of the regular asymptotics already
from the very beginning.

The LS asymptotics has an amazing feature - it allows to
establish the universal form of the size spectrum. One
can admit that the power of metastability has some
universal asymptotics, but one can hardly believe that
the spectrum of the embryos sizes is universal. This
will be the matter of discussion is this paper. Below
it will be shown how the real spectrum of sizes
approaches  the form  given by LS theory.

\section{Some remarks initiated by technique
of Lifshic and Slezov}

Now we shall consider the simplest model leading to the
over-condensation. At first we shall consider the
embryos with linear sizes which strongly exceed the
critical size. We shall call them as the supercritical
embryos. The growth of the supercritical embryo in time
$t$ is given by the following equation
$$
\frac{d\rho}{dt} = \frac{\zeta}{t_1}
$$
Here $\rho$ is the cubic root of the number of
molecules inside the embryo (it plays the role of the
linear size), $\zeta$ is the supersaturation of the
metastable phase, $t_1$ is some characteristic time
(this is simply the constant coefficient). The
supersaturation is defined as the ratio of the surplus
density of the metastable phase (with respect to the
saturated phase) to the density of the saturated phase
or the ratio of the real density of metastable phase to
the density of the saturated phase minus one.

The simplest asymptotic correction to the asymptotic
law of growth will be the following
$$
\frac{d\rho}{dt} = \frac{\zeta}{t_1}(1-u^{-1})
$$
Here $$u=\frac{\rho}{\rho_c}$$ is the ratio of $\rho$ to the critical size
$\rho_c$ to have the zero rate of growth for the
critical embryo. It is quite possible that this ratio
has been changed by an analogous ratio in some power
but this does not lead to the essential reconsideration
of this
approach.

One can give the interpretation of the last correction
term as the leading term in the asymptotic expansion in
 inverse powers of the linear size. It is possible
that this decomposition starts from the non-integer
power as it takes place in construction of series for
solutions of the linear second order differential
equations.
 It is also possible that the law of growth for the supercritical
 embryos has some power of the linear size which grows with a
 constant velocity. Namely this situation takes place in
 the diffusion regime of the metastable phase exchange.
In both cases the technique described below can be
applied.

So, since one can see that
$$
\frac{(d\rho/dt)}{(d\rho/dt)_{asymp}} = (1-1/u)
$$
or there is some function of $u$ in the rhs it is
convenient to consider
 $du/dt$
instead of $d\rho/dt$. For the derivative $du/dt$ we
get an additional term
$$
\frac{d\rho}{dt} = \rho_c  \frac{d u}{dt} +
u \frac{d\rho_c}{dt}
$$
which is linear on $u$. So,
$$
\frac{d u}{dt} = \frac{\zeta}{t_1} (1-u^{-1})
- u \frac{d\rho_c}{dt}
$$

In the LS theory there exists a hidden supposition that
$\rho_c$ depends on time droningly. This supposition
was reexamined in approach which predicts the periodic
formation of the tail of the size spectrum and then the
consumption of this tail. But nevertheless this
supposition is rather natural at least at the
asymptotics. If we adopt this supposition then all
values depending on time are the values depending on
the critical size. At least one can consider the
intervals of the monotonous dependence.

The generalization of this law on other regimes of
growth leads to
$$
\frac{d u}{dt} = \phi_1(t) u^{\alpha} (1-u^{-1})
- u \phi_2(t)
$$
The functions $\phi_1$, $\phi_2$ are the functions of
time. The function $\phi_1$ can be excluded by
transition to the rescaled  time $\tau$. The same can
be done with $u^{\alpha}$.

The right hand side of the last equation (let it be
$\Gamma$) can not attain asymptotically the positive
value at some argument - then the substance balance
will be violated.
 The balance equation will be violated also when the rhs
 is negative at all values of argument
$u$. Hence, the unique possibility is to touch the zero
level at the main maximum. Then it is necessary that
the rhs as a function of $u$ has a maximum and then by
the correct choice of $\varphi$ one can put the maximum
to the zero level. The class of functional dependencies
which allow this operation forms the class of
dependencies where the LS technique can be applied at
least formally. The fine unjustified supposition here
is the monotonous approach (in time) of the maximum to
the zero level. It is quite possible that this maximum
oscillates near zero - at some moments is it greater
than zero, at some times it is less than zero. This
possibility is considered in в \cite{pred}. Here we do
not consider the oscillating scenarios.

For the function $\varphi$ we get
$$ \varphi = \gamma_0 $$
where $\gamma_0$ is some constant. Even when this
supposition is adopted the velocity at the maximum is
zero and the balance equation is violated. Then it is
necessary to be:
$$
\varphi = \gamma_0 (1
+ \varepsilon^2(\tau))
$$
with $\varepsilon \rightarrow 0 $ at $\tau \rightarrow
\infty$.

The reason that $\gamma_0$ is a constant lies in the
form of equation for $du/d\tau$. It would be
interesting to consider equations which do not belong
to the established form and to get the evolution
analogous to the over-condensation.

Let the argument of the main maximum be $u_m$. One can
see that at $\varphi = \gamma_0$ the embryos with
$u>u_m$ can be dissolved only at the infinite time (the
diffusion is not considered here).

Now we shall establish the function
$\varepsilon(\tau)$. To find this function one has to
analyze the solution near the maximum of $\Gamma$. We
introduce the variable
$$
z= (u-u_m)/\varepsilon
$$

Then we see that
$$
du/d\tau = d(u-u_m)/d\tau = dz\varepsilon/d\tau
= \varepsilon dz/d\tau + z d\varepsilon /d\tau
$$
and
$$
\varepsilon \frac{dz}{d\tau} =
\Gamma(\gamma_0(1+\varepsilon^2)) -   z k_1
$$
where $k_1$ is the coefficient equal to $d\varepsilon
/d\tau$. In the last equation it is necessary to
express $u$ in the function $\Gamma$ through $z$ and
$\varepsilon$. Then we rescale the time to exclude the
coefficient in the rhs and then we get an equation
which allows (or does not allow according to the
reasons mentioned above) the analysis analogous to the
already made one. If equation allows such analysis one
can get the correction term of the asymptotic.

One has to get the equation binding  $d\varepsilon
/d\tau$ and $\varepsilon$.
In the power-like laws of growth this equation is
trivial and has the form $f(\varepsilon)
 d\varepsilon /d\tau = const$
with some known function $f$. It is interesting whether
one can get any more complex form of this equation. It
is quite possible that some more complex form can lead
to some new physical effects. Here we do not consider
this question and assume that the dependence
$\varepsilon$ on $\tau$ for the correction term is
established.

Having calculated one correction term after another we
establish all asymptotic series until the step when we
can not perform the procedure described above because
the obtained equation does not allow to put maximum to
the zero level. In the natural physical situations only
the initial and the first correction term can be
established. It is worth seeking the situations where
one can make more steps.

In the Osipov-Kukushkin approach we get the correction
term already as the initial approximation since one can
choose such variables where the correction term can be
treated as the initial one. Certainly, it is possible to
continue this procedure but it appears that under the
natural regimes of the substance exchange it is
impossible to choose parameters in equations for
correction terms that the main maximum
will touch the zero level. It is interesting to find
situations where all corrections are the universal
ones.

One can see that the account of the diffusional term in the
evolution equation can not lead to another asymptotics
because the diffusion process along $u$ becomes
negligible at the infinitely big time. In correction
terms it is necessary to check that the diffusion
process is really negligible in account of corrections.
It would be interesting to seek the situations where
the account of the diffusion process changes the
correction terms.

The unpleasant fact for the LS theory is the following
one. In the zero approximation the time of the
dissolution of the embryo with
 $u>u_m$
is infinite and has to be refined. But in the first
approximation the time of dissolution of every finite
embryo is the finite one and it means that every finite
size spectrum will be dissolved (dissolved at the
finite time). After the spectrum is dissolved the
balance equation will be certainly violated.

The mentioned difficulty has a fundamental character
which is confirmed by further constructions in frames
of the LS theory. It is reasonable to connect the
finite time of dissolution with the possibility to find
the universal form of the size spectrum. The authors of
the LS approach treat the universal distribution as the
 distribution which is the limit of the
relaxation process. But it evident that namely the
initial form of the size spectrum determines the whole
further evolution. Under the regular and only under the
regular law of growth we have
$$
p(\rho, t) d\rho = f(\rho', t') d\rho'
$$
where $p$ and $f$ are the old spectrum and the new
spectrum of sizes and $\rho$ at the moment $t$ has to
come by the regular growth into $\rho'$ at $t'$. Then
$$
p(\rho, t) \frac{d\rho}{dt}
= f(\rho', t') \frac{d\rho}{dt}|_{t=t',\rho=\rho'}
$$
Since for $d\rho/dt$ we have the concrete given
expression it is hardly possible to change the form of
the size spectrum to come to the universal form of the
LS theory. If one determines the velocity of growth
from comparison of the initial form of spectrum and the
final universal form of spectrum it leads to the
unpleasant contradiction.

The alternative is the following: to have the size
spectrum
prescribed by the initial distribution or to see the
leading role of diffusion (at least at some time).

It appears that the universal form of the size spectrum
has absolutely another sense - this form of spectrum is
such a form which corresponds to the already
established asymptotics for the critical size precisely
or at least ensures the optimal relaxation to the
established asymptotics for the velocity of growth. With
the real form of the size spectrum this asymptotic form
of the size spectrum has no direct connection. Then the
asymptotic for the velocity of growth corresponding to
the ideal size spectrum will be never attained at
finite time.

\section{Asymptotic form of the size spectrum}

One can get the universal form of the size spectrum in
the LS formalism rather simply. In the LS approach it
is supposed that the main quantity of the substance is
in the region $u<u_m$. Then the balance of the
substance leads to
$$
\int_0^{u_m \rho_c}\rho^3 f(\rho, t) d\rho = const
$$
where $f(\rho, t)$ is the distribution function. From
this function we come to the distribution over $u$,
namely to $\phi(u, \tau)$. Then
$$
 \phi(u, \tau) du =  f(\rho, t) d\rho
$$
The substance balance can be rewritten as
$$
\rho_c^3 \int_0^{u_m} u^3 \phi(u,\tau) du = const
$$
The form of the function $\phi$ can be determined from
the continuity equation
$$
\frac{\partial \phi}{\partial \tau} =
- \frac{\partial \phi v_u}{\partial u}
$$
where $v_u$ is the velocity of the growth for $u$. The
solution of this equation is rather simple
$$
\phi(u,\tau) = \theta(\tau - \tau(u))/v_u
$$
where $\tau(u)$ is the time for the embryos to attain
$u$. Instead of the last function one can write the
function of $u - u(\tau)$, where $u(\tau)$ is the size
attained at the given time. The sense of solution is
that the dependence over two variables is performed
through the dependence over one variable. The concrete
form of this dependence is determined by initial
conditions (in the
LS approach it is detemined by asymptotic relations).

The transformation to the variable $u$ is made to
ensure the constant value of the upper limit of
integration.

Ordinary the concrete form of the functional dependence
of $\theta$ has to be determined from the initial
conditions. But in the LS formalism this form is
determined from the asymptotic balance equation which
can be written as
$$
\rho_c^3 \int_0^{u_m} u^3 \theta(\tau - \tau(u))/v_u du = const
$$
The function $v_u$ depends only on $u$, but not on
$\tau$. This is the consequence of the fact that in the
zero approximation for $du/d\tau$ one can take
$\gamma_0$ which excludes the dependence on time.

The function $\rho_c^3$ at the already established
asymptotic is the known function of time. This
dependence has to be cancelled by the true choice of the
functional dependence for $\theta$. Namely this
cancellation is the recipe to choose the true form of
the function $\theta$.

The simple dependence (but may be not the unique one)
is the following one
$$
\theta(\tau-\tau(u)) = g(\exp(\tau-\tau(u)))
$$
where the function $g$ satisfies the relation $$g(ab) =
g_1(a) g_2(b)$$ for any $a$ и $b$. As this function one
can take the power function. Since we have to cancel
only the leading term in asymptotic the choice of the
power-like function is quite satisfactory.

So, the solution is announced. But is it well
justified?

Really, the ordinary solution of the problem (not the
asymptotic one) is very simple – it is already
presented in the form of the function $\theta$, i.e. in the
fact that $\theta$ is the function of only one variable.
Certainly, this function has to coincide
with the initial spectrum
of sizes at the initial moment of time. It is
sufficient to calculate the origin of the given embryo
taking into account the known supersaturation, then to
take the initial distribution and divide it on the
velocity of growth in the given point (or on the ratio
of the velocities of growth in corresponding points).

We shall speak here about the evolution scenario. This
scenario is in contradiction with the LS scenario. It
conserves the explicit dependence on the initial
distribution. It is clear that the evolution scenario
is more correct in comparison with the LS scenario. The
LS scenario has at least two disadvantages
\begin{itemize}
\item
-The absence of the spectrum at  $u>u_m$

\item
-The spectrum is determined to ensure the asymptotic,
but it is more reasonable to get the asymptotic on the
base of spectrum.
\end{itemize}

The first disadvantage can be ignored by notation that
every finite spectrum sooner or later will leave the
region $u>u_m$.

The second disadvantage is more serious because it
destroys the methodology of the LS approach. It seems
that it is the crucial point and the real size
spectrum will differ from the result of the LS theory.

But one can see the amazing fact - the similarity
between the form of the LS spectrum and the results of
experiment. We shall explain this similarity below.

\section{Approximate similarity between the real size
spectrum and the results of the LS approach}

We start to consider the strange fact - why the
form of the size spectrum prescribed by initial
conditions will resemble the universal result of the LS
theory?

At first one has to stress the low accuracy in the
experimental determination of the size spectrum form.
As an example one can consider the style of discussion
in \cite{kuk}. In \cite{kuk} one can see not only the account
of the initial approximation but already of
the first correction in the form of the size spectrum.
The form of the size spectrum with the first correction
essentially differs from
the form in the zero approximation. This fact is very
important and leads at least to two essential
conclusions
\begin{itemize}
\item
- the correction term in the asymptotic expansion is not
small at least for the size spectrum - one can not see
the parameter with a property: the small value
of this parameter leads to the size
spectrum in the zero approximation. Moreover, one can
state that there is no such parameter because the size
spectrum in the first approximation is universal one.
This shows that there is no reliable way to determine
the size spectrum because the next correction can
change the form of the size spectrum radically.

One can also give an interpretation which is not
favorable for the
LS approach - the modified zero (Osipov-Kukushkin)
approximation is only the starting point and all
further terms of decomposition depend on the initial
size spectrum. So, there is no reason to speak about
the universality.

\item
- the accuracy of the experimental results is rather low.
Really, the difference between the size spectrum in the
zero (LS theory) and in the first approximation is very
essential. Earlier the experimental results confirmed
the LS spectrum in the zero approximation. Now
experiments confirm the Osipov-Kukushkin result. It can
be only when the accuracy is low. So, one can speak
only about the experimental observation of some
tendencies in the form of the size spectrum.

In the analysis of experimental results one has to note
that some authors speak about the diffusional blurring
of the size spectrum which corresponds to experiment.

But one can show analytically that it is possible to
neglect the diffusional term in the LS technique. It
does not mean that in the evolution scenario one can
neglect the diffusion - at least there are some periods
when the diffusion is the driving force of evolution.
Also one can see that in the case of the finite size
spectrum one has to include diffusion.
\end{itemize}

As the result of these facts one has to conclude that
there exists a tail of the size spectrum (it is the
exponential one - this is explained by the
diffusion process) at $u>u_m$ (in any other region it
can not be noticed). This tail is observed
experimentally. What is the reason for this tail? The
LS formalism can not give an answer. Below this answer
will be given.

Kinetics of the new phase embryos formation has some
characteristic features which help to determine the
characteristic features of the over-condensation. Roughly
speaking,
the "initial" size spectrum belongs to a narrow class
of functional dependencies.  So,
we use the evolution approach and construct the
sequence of stages for the process of condensation and
over-condensation.

The process of condensation (the periods before the
over-condensation) is investigated in \cite{inst}. This
investigation gave the total number of the embryos and
the form of the size spectrum. For the further regular
evolution it is necessary to know the first three (and
the zero) momenta of the size spectrum
$$
\mu_i = \int_{-\infty}^\infty f(x) x^i dx
$$
or of the distribution function $f(x)$ of the variable
$x$, defined as the deviation of the coordinate $\rho$
from the size $z$ of the maximal value of this variable
corresponding to the embryo formed at the very
beginning of the condensation process.

\subsection{The period of the
initial relaxation}

The relatively intensive formation of droplets stops at
the relative decrease of the supersaturation equal to
the value reciprocal to the quantity of the molecules
in the critical embryo.

The number of molecules in the critical value is a big
value. The relatively small decrease of the
supersaturation stops the nucleation (formation of new
embryos) and later the size spectrum moves along the
$\rho$-axis without any change of the form. To ensure
the essential exhaustion of the metastable phase the
spectrum has to move along $\rho$-axis for a rather
long distance. So, at the end of this evolution the
spectrum of sizes can be considered as the monodisperse
one.

The balance equation can be written in the following
form
$$
\Phi =
\zeta + \sum_{j=0}^3 \frac{3!|}{i!(3-i)!} z^{(3-i)} \mu_i
$$
or
$$
\Phi = t_1 \frac{dz}{dt}
 + \sum_{j=0}^3 \frac{3!|}{i!(3-i)!} z^{(3-i)} \mu_i
$$
which is the ordinary first order differential equation
without the explicit dependence on the argument. So, it
can be easily integrated. The result is the
monodisperse spectrum and the relaxation of the
spectrum coordinate to the critical size (or more
correct the relaxation of the critical size to the
coordinate of the spectrum).

Here we use the law of growth for the supercritical
embryos. It can be replaced by the precise law of
embryos growth. Really, in the law
$$
\frac{dz}{dt} = \frac{\zeta}{t_1} ( 1- \frac{z 3
\zeta}{2a})
$$
we replace $\zeta$ by $\Phi - \sum_{j=0}^3 \frac{3!|}{i!(3-i)!} z^{(3-i)} \mu_i
$ and get equation
$$
\frac{dz}{dt} = \frac{\Phi - \sum_{j=0}^3 \frac{3!|}{i!(3-i)!} z^{(3-i)} \mu_i
}{t_1} ( 1- \frac{z 3
(\Phi - \sum_{j=0}^3 \frac{3!|}{i!(3-i)!} z^{(3-i)} \mu_i
)}{2a})
$$
which can be easily integrated.
Here it is ignored that during the evolution
the momenta $\mu_i$ will be changed which is considered
as a correction. In any case we need only the initial
approach to the critical size where the rough
monodisperse approximation is sufficient. Then there
will be no problem with changing momenta.

One can show that the account of the diffusional term
will be essential only when the coordinate of the size
spectrum is rather close to the critical size. This is
the end of the relaxation stage and the beginning of
the new period.

\subsection{The period of
the diffusional blurring of the size spectrum}

The result of the previous period is the relaxation of
the spectrum coordinate to the critical size. The
spectrum resembles the delta-like function, The
spectrum width $\delta
\rho$
is many times less than the coordinate $\rho$ or
$\rho_c$ which is the spectrum coordinate.

If there would be no diffusion then the spectrum will
remain near the critical coordinate until the end of
the whole evolution. But after the time of relaxation
at the previous period the diffusion becomes the main
driving force of the process.

Kinetics of the diffusion blurring is rather simple and
it is described in \cite{pred}. It is possible to
approximate evolution by diffusion blurring without any
regular growth with a boundary condition
$$
f(\rho=0) = 0
$$
and the initial condition
$$
f(t=t_{initial}) \sim \delta(\rho-\rho_c)
$$

The method to solve this problem is the combination of
the Green functions at the infinite interval. The
method of images allows to construct solution by
addition of the negative gaussian in the symmetrical
point.

So, we write the diffusional equation in the following
form
$$
\frac{\partial f}{\partial t} = W_c
 \frac{\partial^2 f}{\partial \rho^2}
$$
Here $W = W^+ + W^-$ is the generalized kinetic
coefficient equal to the weighted sum of the adsorption
coefficient $W^+$ and the ejection coefficient $W^-$.
One can take these coefficients in the critical point
marked by the index $c$. Then one can assume that $W_c
= 2 W_c^+$. Taking into account the evident relation
$$
W^+= Sv_t/4 = 3 \rho^2 (\zeta+1)/t_1
$$
 where $v_t$ is the mean
thermal velocity, $S$ is the surface square of the
embryo one can determine the dependence of $W_c$ on
$\rho$. Then it is necessary to go from $\rho$ to a new
variable $r$ which is $\rho$ in some constant power,
i.e. $\rho^{const}$. This transition excludes the
dependence of the diffusion coefficient on the size at
least asymptotically. It occurs at $ \rho d \rho \sim
ds = d\rho^2 $.

The Green function at the infinite interval is written
in the following form
$$
G(s,t|s_0,t_0) \sim
\exp(- \frac{(s-s_0)^2}{4 D (t-t_0)} )
$$
where $D$ is the diffusion coefficient (this is the
known
constant value), $s_0$ is the point of appearance of
elementary disturbance at the moment $t_0$, $s$ is the
point of observation at the moment $t$.

Here one can take as $s_0$ the critical size and the
time $t_0$ has to correspond to the time of the end of
relaxation (actually it is the time of relaxation).

To observe the boundary condition $f(s=0)=0$ it is
necessary to take the combination
$$
f_0 = f_+ + f_-, \ \ f_+ =G(s,t|s_0,t_0),
\ \ f_- = G(s,t|-s_0,t_0)
$$
This gives the solution of this problem.

Consider the behavior of the critical size $\rho_c$.
It is important to know $\rho_c$ because $u=\rho/\rho_c$.
 One can propose the equilibrium critical size
 $\rho_{ce}$ which can be determined on the base of
 the size spectrum as
 $$
 \int_{-\infty}^{\infty} f(\rho,t) 3 \rho^2
 (1-\frac{\rho_c}{\rho}) d\rho = 0
 $$
This corresponds the stationary value of the
critical size, i.e. $d\rho_c/dt =0$.

It is clear that never $\rho_c$ equals $\rho_{ce}$
because this means the stationary value of $\rho_c$ and
of the supersaturation $\zeta$. But approximation
$\rho_c \approx \rho_{ce}$ is rather good. Namely, at
the beginning of the diffusion blurring this equality
takes place. So,
 the critical radius $\rho_c$ (we shall
mark it $\rho_{c0}$ for initial time) is given by
condition
$$
\int_0^\infty f_0(\rho) v_\rho(\rho_{c0}) dr =0
$$
Here $v_\rho$ is the velocity of growth of the variable
$\rho$. It is supposed that the size spectrum is
relatively narrow. Another variant taking into account
the different volumes of embryos is the following
$$
\int_0^\infty f_0(\rho)  \rho^2 v_\rho(\rho_{c0}) dr =0
$$

Also
one can propose to extract the deviation of $\rho_c$
from $\rho_{ce}$
$$ y=(\rho_c-\rho_{ce})/\rho_{ce}
$$
and see that ordinary $y$ is small. Then one can
analyze the evolution of the system through
decomposition on $y$.

When $y$ is big it means that the size spectrum is
essentially greater than $\rho_c$. But this
corresponds to the evolution via supercritical embryos
where we have extremely simple law of growth $d\rho/dt
= \zeta/t_1$. So, the combination of the consideration
of supercritical embryos and decompositions on $y$
with restriction in the first several terms (actually the
first non-zero term) can be very effective.

Now we return to the diffusion blurring.
The result of the diffusion blurring is very optimistic
for the final conclusions. It sounds as following:
The distribution function is the universal one and does
not undergo the change of the form
any more. So, one can see that the
asymptotic solution is found. But the situation is not
so simple.

Really the function $f_0$ after scaling in units of
$\rho_{ce}$ will be the universal function without any
parameters.

On the base of distribution we can calculate the
behavior of the critical radius. Note that it is
impossible to find the critical size directly from the
balance equation $2a/3\rho_c + \int_0^\infty
\rho^3 f(\rho, t) d\rho = const$ because here
$\int_0^\infty
\rho^3 f(\rho, t) d\rho \approx const$ and the error
radically increases. Instead of the direct balance
equation one can use the differentiated variant
$2(a/3\rho_c^2) (d\rho_c/dt) =  3 \int_0^\infty
\rho^2 f(\rho, t) d\rho$
which allows to find $\rho_c$

If the size spectrum will be the universal function
then the critical size will be also the universal
function.

The special question is the correct boundary condition
at small sizes. The velocity of the dissolution of the
small embryos is a rather complex function of size and
the regular dissolution exists. One can not neglect
this regular dissolution. But fortunately the small
embryos are dissolved very quickly with the growing
velocity. So, one can suppose that they disappear
immediately at $\rho=\rho_f=(0.6 \div 0.8) \rho_c$.
So, the zero
boundary condition has to be put at $\rho_f$ and all
further considerations remain without reconsideration.
Certainly, to keep the boundary condition we have to
put the negative Green function symmetrical to $s$ with
respect to $s_f$.

In the situation with $s_f$ we have also the universal
spectrum and the universal behavior of the critical
size.

But this universal asymptotics is only the
intermediate asymptotics.

Now we shall introduce the regular growth and destroy
this universality.

It is necessary to put some boundary of the type
$$
\rho_r = (2 \div 3) \rho_c
$$
and for the sizes greater than $\rho_r$ one has to
consider the regular motion
with the asymptotic velocity. The choice of $\rho_r$ can
be made also on the base of the LS analysis.

It is trivial to refine the
solution by investigation of the transition zone
explicitly.

The growth of the supercritical tail leads to the
growth of the size $\rho_c$ which can be calculated in
 approximation of the following iteration
procedure: On the base of initial $\rho_{c0}$ we find the
tail $f_{tail}$ of the size spectrum
$$
f_{tail}(\rho,t) = f_0 ( \rho_{r}, t') v_\rho( \rho_{r})/
v_\rho (\rho)
$$
where
$$
t-t'=\int_{\rho_{r}}^\rho \frac{1}{v_\rho(\rho')} d\rho'
$$
(the explicit dependence of $v_\rho$
on $t$ is weak or it can be
expressed via $t$ iteratively).

The new distribution function $f_1$
will be the superposition of the initial part $f_0$ and the
tail. Then from the balance equation
$$
\int_0^\infty f_1(\rho)  \rho^2 v_\rho(\rho|\rho_{c1})
d\rho =0
$$
we find $\rho_{ce1}$. This will be the new equilibrium
critical size. It will be near the real critical size
unless the tail begins to play the main role in the
metastable phase consumption.

The period of the diffusion blurring come to the end
when the velocity of the growth for the critical embryo
becomes to be comparable with the velocity of the
growth for the tail.

One has to analyze an attractive possibility to
consider the process of diffusion in the region
$s<s_r$
with the linear on
 $s-s_c$ rate of growth. Here $s_r$
has to be determined as the boundary between the linear
and the asymptotic rates of the embryos growth. It
seems from the first point of view that the linear rate
of growth in the near critical region is preferable
instead of the absence of the regular growth considered
above.

The Fokker-Planck equation under the linear rate of
growth has the following form
$$
\frac{\partial p}{\partial t} =
\gamma \frac{\partial yp}{\partial y} + D \frac{\partial^2
p}{\partial y^2}
$$
Here the zero value of the coordinate as the critical size
is taken, $\gamma$ and $D$ are some constants. Here
$\gamma$ is negative. The Green function for the
positive $\gamma$ at the infinite interval is well
known and has the following form
$$
G (x,t|x',t') =
\sqrt{\frac{\gamma}{2 \pi
D (1-e^{-2\gamma (t-t')})}}
\exp(- \frac{\gamma(x-e^{-\gamma(t-t')}x')^2}{2D
(1-e^{-2\gamma(t-t')})})
$$
Now it is necessary to take the combination of two
Green functions and the answer is ready. The further
analysis is absolutely the same.

This approach seems to be more precise than the
previous one but it has many disadvantages. The first
disadvantage is the following: one can see that here
$\gamma$ has to be negative and then at some time the
half width of the gaussian goes to infinity. So, the
solution becomes illegal.

The second disadvantage is how to take into account the
drift of the critical size. Now it appears in the rate
of growth and then in the final formulas. The solution
with a moving critical size is illegal also.

But the idea to consider the law of growth as a
combination of the linear dependence at $\rho<\rho_r$ and the
asymptotic law at $\rho_>\rho_r$ is very attractive. Really
the term $1-u^{-1}$ in the traditional law of growth
can be treated as a correction term in the asymptotic
decomposition. Here this asymptotic is taken over the
positive powers of $\rho_r$. The decomposition on the
positive powers is not less justified in comparison
with the decomposition on inverse powers. But the last
approximate rate of growth allows an explicit
integration and then the LS technique can be analyzed
explicitly.

One can try to construct the approximate Green
function for the case of the presence of the  regular
growth in the following manner. We construct this
function for initial perturbation at $x_0=0$ which is
an equilibrium value for the regular growth
$v_x(x=0)=0$. We suppose that there are no other
points where $v_x=0$ and $v_x(x)=-v_x(-x)$. Then we
seek the Green function in the ordinary form
$$
G(x,t|t_0)= A(t,t_0)
\exp(-\frac{x^2}{\Delta(t,t_0)^2})
$$
where the amplitude $A$ can be reconstructed on the
base of normalizing equation $\int G dx =1$ and the
width $\Delta$ is found by relation
$$
\Delta = \sqrt{4D(t-t_0)} + \int_{t_0}^t v_x (\Delta(t')
) dt'
$$
or
$$
\Delta = \sqrt{4\int_{t_0}^t D(\Delta(t'))
 dt' } + \int_{t_0}^t v_x (\Delta(t')
) dt'
$$
for varying $D$. The last equation can be easily
solved  iteratively
$$
\Delta_{i+1} = \sqrt{4\int_{t_0}^t D(\Delta_i(t'))
 dt' } + \int_{t_0}^t v_x (\Delta_i(t')
) dt'
$$
$$
\Delta_0 = \Delta(t_0)
$$

To ensure the correct boundary condition it is
necessary to add the symmetrical negative Green
function.

To refine the solution one can also use here the values of
effective diffusion coefficient and effective law of
the regular growth velocity from consideration made in
\cite{stohgr}.

\subsection{The period of the dissolution of the head
of the size spectrum}

The tail of the size spectrum grows and earlier or
later the main role of the metastable phase
consumption will belong to the tail. This opens the
period of dissolution of the head of the spectrum.

This period allows a rather trivial description since the high
accuracy is not important here. Inevitably the head of the spectrum
will be dissolved and this marks the end of this period.

The most primitive description is the following. We
split the substance between the tail $G_{tail}$ and
the head $G_{head}$. The spectrum in the head is
described by $f_0$. The spectrum in the tail
$f_{tail}$ is the direct translation of the blurring
part of the head which comes to the zone $\rho>\rho_r$
by the regular growth with
the known
supersaturation. The supersaturation
 corresponds in the first iteration loop
 to the stationary
position of the critical size. Then we can calculate
the dissolution of the head, the growth of the tail
and replace the critical size on the base of the
balance equation.  We know a new value
of supersaturation. This closes the
iteration loop.

Here it is impossible to use the model of the growth
with a zero value of growth for $\rho>\rho_f$ and the
zero value of $\rho$ for all $\rho<\rho_f$ (otherwise
it produces the jumps in the supersaturation value).
One has to use the
explicit law $d \rho / dt =
(2a/3\rho_c)(1-(\rho_c/\rho))$.

Another style of description is to use the methods
from description of the dissolution of the tail of the
size spectrum which is analyzed below. Since the
method is the same we do not consider it here
explicitly. Certainly, the exponential tail like
$\exp(-const \ \rho)$ has to be changed to the head of
the size spectrum $f_0$.

This period is rather short and it is not very
important for the further evolution.
Details of this period description can be found in
\cite{pred}. But one can see that in the theory
presented here the new head at the tail
is not formed. Here lies the main difference between
this theory and the theory from \cite{pred}. The
question whether the new head is formed is rather
complex. This is the question of applicability of the
gaussian tails of the Green function of diffusion
equation. If we adopt the model with a finite
 upper limit of the size spectrum then we have a new
 head at the tail and have to use scenarios proposed
 in \cite{pred}. If we believe in long gaussian tails
 we come to the theory presented here. To solve this
 question we must go ahead of the level of description
taken in the diffusion approximation. Otherwise there
is no sufficient statistics to solve this question. We
prefer to stop here at the statement that there is no
sufficient statistics. It means that concrete details
will determine the scenario. For example, we adopt
that the act of the molecule consumption by the embryo
requires a certain elementary time and we come to the
finite upper limit of the spectrum. In the opposite
situation we come to the gaussian tail. Certainly this
question is out of the level of consideration adopted
in the nucleation theory.

\subsection{The period of the gradual consumption of the tail}

Now we come to the period which in some sense resembles
the LS results.

The result of the previous period is the formation of
the exponential tail at $\rho>\rho_r$. Now we consider the
process of the tail dissolution.
Here one can see the
true competition between the embryos with different
sizes. So, here it is convenient to go to the LS
coordinates.

To investigate this period we simplify the rate of the
embryos growth. We assume the following rate of growth
$$
\frac{du}{d\tau} \sim (1-u^{-1}) - \gamma u
$$

If there is a sufficient tail and the heat is already
dissolved then at the moment of the end of the
previous period
$$
max\{ \frac{du}{d\tau} \} \equiv v_m < 0
$$
So, we assume that $v_m$ is negative.

We split the whole region of $u$ into three small
regions. In the region of big $u$ we suppose
$$
\frac{du}{d\tau} \sim (1-0) - \gamma u
$$
This law allows integration even under the variation of
$\gamma$.

The next region is the region of the intermediate $u$.
Here we assume
$$
\frac{du}{d\tau} = v_m
$$

In the region of small $u$ we neglect $\gamma u$ and
get
$$
\frac{du}{d\tau} \sim (1-u^{-1})
$$
This law does not contain parameters and can be easily
integrated. It means that the dissolution here is so
fast that we can neglect the change of the critical
size.

The boundaries $u_1$ and $u_2$ between these regions
can be established from the continuity of the rate of
growth.

One can also put an effective boundary of the total
dissolution instead of the zero size.

It is possibly to refine the law of growth having
introduced instead of $v_m$ some other effective value
of the flat region.

When the size spectrum is known then the balance of the
substance becomes the transcendental equation on
$\gamma$. After we found $\gamma$ we can get $\rho_c$
by integration.

One can easily follow the dissolution of the spectrum
on the base of the approximate rate of growth.

We accumulate the approximate law of growth in the
following formula
$$
du/d\tau \approx (du/d\tau)_{appr}
$$
This law of of growth allows to know $u(\tau)$ on the
base of some $u_0(\tau_0)$ for every arbitrary curve
$\gamma(\tau')$
$$
u(\tau) = u_0(\tau_0) + \int_{\tau_0}^{\tau}
(du/d\tau)_{appr}
\equiv F_{appr}(\tau | u_0, \tau_0 ; \gamma(\tau'))
$$
For the law of growth we write
$$
du/d\tau = \varphi(u, \gamma(\tau))
$$
Here we can take both approximate or precise law of
growth.

The substance balance equation
\begin{eqnarray}
\nonumber
\frac{d\rho_c}{d\tau} \frac{2a}{3 \rho_c^2(\tau)}
=
\rho_c^3(\tau)
\int_0^{\infty} 3 F^2_{appr}(\tau | u_0,\tau_0;d\ln
\rho_c(\tau')/d\tau')
\\
\varphi(F_{appr}(\tau | u_0,\tau_0;d\ln
\rho_c(\tau')/d\tau'),d\ln\rho_c/d\tau)
f_0(u_0, \tau_0) du_0
\nonumber
\\
+3 \rho_c^2(\tau) \frac{d\rho_c}{d\tau}
\int_0^{\infty}  F^3_{appr}(\tau | u_0,\tau_0;d\ln
\rho_c(\tau')/d\tau')
f_0(u_0, \tau_0) du_0
\nonumber
\end{eqnarray}
is now the closed equation on $\rho_c(\tau)$.
Here instead of $a$ one can put the appropriate
constant in accordance of normalization of the size
spectrum.

It is necessary
to stress that all functional dependencies here are explicit
ones and except $\rho_c(\tau)$ all other dependencies are
known. The best way to solve this equation is to use
the steepest descent method.
The methods
to solve this equation will be discussed below.

This equation can be approximately simplified. Since
the last term of the rhs is many times greater than
the lhs one can approximately write
\begin{eqnarray}
\nonumber
\rho_c(\tau)
\int_0^{\infty} 3 F^2_{appr}(\tau | u_0,\tau_0;d\ln
\rho_c(\tau')/d\tau')
\\
\varphi(F_{appr}(\tau | u_0,\tau_0;d\ln
\rho_c(\tau')/d\tau'),d\ln\rho_c/d\tau)
f_0(u_0, \tau_0) du_0
\nonumber =
\\
-3  \frac{d\rho_c}{d\tau}
\int_0^{\infty}  F^3_{appr}(\tau | u_0,\tau_0;d\ln
\rho_c(\tau')/d\tau')
f_0(u_0, \tau_0) du_0
\nonumber
\end{eqnarray}
or
\begin{eqnarray}
\nonumber
\frac{
\int_0^{\infty} 3 F^2_{appr}(\tau | u_0,\tau_0;d\ln
\rho_c(\tau')/d\tau')
\varphi(F_{appr}(\tau | u_0,\tau_0;d\ln
\rho_c(\tau')/d\tau'),d\ln\rho_c/d\tau)
f_0(u_0, \tau_0) du_0}
{ 3\int_0^{\infty}  F^3_{appr}(\tau | u_0,\tau_0;d\ln
\rho_c(\tau')/d\tau')
f_0(u_0, \tau_0) du_0
}
\nonumber \\ = \nonumber
- \frac{d\ln\rho_c}{d\tau} \equiv - \gamma
\end{eqnarray}

This equation can be solved by the same methods but it
is more simple than the previous one. We outline again
that except $\gamma(\tau)$ all other dependencies here
are known.

Another possible approximate variant of the balance
equation is the following

$$
\rho_c^{-3} = const^{-1}
\int_0^{\infty}  F^3_{appr}(\tau | u_0,\tau_0;d\ln
\rho_c(\tau')/d\tau')
f_0(u_0, \tau_0) du_0
$$
It seems to be the most simple variant of the balance
equation.

Now we shall discuss the asymptotic properties of the
balance
equation.

Generally speaking the problem is solved since we know
the good approximation
$$
d\ln \rho_c / dt
= \gamma_0 + some\ \ positive\ \ small\ \ value.
$$
We can solve this equation by decomposition in series
or by some effective linearizations.

But below we
shall analyze the properties of solution in order to
see that the size spectrum here resembles the LS theory
for the size distribution.

First of all we have to note that the tail has the
exponential character. Really, the translation of the
gaussian at some shift from the maximum leads to
$$
f_{tail} \sim \exp (- const/t)
$$
which can be easily approximated by the standard
exponent of the argument linear on the size. Here the
$const$ is some fixed value proportional to
$(s_r-s_c)^2$.

Here we have to recall that in the original paper by
Lifshic and Slezov there is a reference on the
exponential tail of the size spectrum. It is quite
natural to check the theory on example of the
exponential tail.

The exponential on $r$ spectrum is exponential on $u$
also in the asymptotic limit.

In frame of the steepest descent methods the
utilization of the exponential tails is quite
justified. One can simply refer to the steepest descent
method instead of the explicit consideration made
above. But one has to stress that we follow the
explicit determination of the size spectrum instead of
the formal methods. Explicit decompositions also give
the exponential tail of the spectrum.

If the characteristic width of the tail is many times
greater than the critical size then $v_m$ is far from
zero and there appears the rapid dissolution of the
size spectrum. If the width of the spectrum is many
times less than the critical size then $v_m$
is close to zero. It is evident that earlier or later
the last
situation will take place. One can give the
qualitative picture
of the process -
The evolution at the big finite time is the slow
monotonous increase of $v_m$ up to zero.

The situation of the wide tail can be investigated
rather elementary. The behavior of supersaturation is
governed by the consumption of the substance by the
wide tail. To see this consumption one can use the
regular growth. This is described in \cite{pred} under
the investigation of the oscillating regime. Evidently,
the consumption of the substance leads to the growth of
the critical embryo and the dissolution of the tail.
This process will take place until the tail (or the
rest of the tail) can be considered as the wide one.

The rest of the tail earlier or later will become the
narrow tail and then we can use the theory of the
narrow tail.

Now we consider the situation of the narrow tail.

We consider the form of the size spectrum. We use the
known formula
$$
f(u,\tau) \sim - \frac{\xi(\tau - \tau(u))}{v_u}
$$
where $v_u$ is the velocity in $u$-axis,
$\tau(u)$ is the time for the embryo to  attain $u$.

One can note that
$$
\tau(u) = \int_0^u \frac{du}{v_u} \rightarrow
\ln(u)
$$
This asymptotics makes the size $u$ inconvenient for
analysis. It is more convenient to act in the
$\rho$-scale where the asymptotic rate of growth is the
constant one. In experiment under the instantaneous
observation the variable $u$ is proportional to $\rho$
and there is practically no difference between them.

So, it is preferable to consider at big $u$ the
$\rho$-scale. In the variable $\rho$ the picture is
rather simple - the exponential tail begins to be
transformed according to the variation in the velocity
of growth
$$
f(\rho) \sim \frac{\exp( - const \rho)}{(1-\rho_c/\rho)}
$$

The amplitude of the spectrum and the width are
determined from the behavior of $\rho_c(t)$. Because
of the asymptotic neighborhood of the behavior of the
critical size to its behavior is the LS model these
characteristics are close to the results of the LS
technique.

Now we turn to the justification of the similarity of
the form of the size spectrum in the LS technique and
the spectrum established in this theory.
Until $u \approx u_m$
or $\rho_0 \approx u_m
\rho_c$
there is no spectrum in the LS theory. In the current
theory the situation is analogous - the tail is very
short.

Now we investigate the region $u \approx u_m$. Since
$u_m$ is big the rate of growth $d \rho
/ dt$
is close to the asymptotic value, i.e. to the constant
velocity and the tail in the current model will be
close to the exponential one. But what will be in the
LS theory? We turn to the formula
$$
\theta(\tau-\tau(u)) = g(\exp(\tau-\tau(u)))
$$
which can be rewritten with account of an initial form
of the size spectrum as
$$
\theta(\tau-\tau(u)) = \exp(const(\tau-\tau(u)))
$$
Having recall that
$$
\tau (u) = - \int_0^u \frac{du}{v_u} + const
$$
which gives under the constant value of $v_u$ the
evident relation near the maximum
$$
\tau(u) \sim u + const
$$
we see that the dependence of $\theta$ on $u$ ($\tau$ is
fixed but it is excluded) becomes the exponential one.

Certainly here the derivation differs from the LS
analysis and we ignore the change of $v_m$ in time
which can be very essential. But qualitatively we come
to the same results.

The distributions in the region with small $u$ are
formed both in the LS theory and here by the
dissolution of the exponential spectrum. They are,
hence, similar.

When one neglects $\gamma u$ in comparison with
$(1-u^{-1})$ it means that we neglect the change of the
height of the original spectrum because the time of the
dissolution of the given embryo from the size $u \sim
u_m$ is small and the change of the critical size
during this time is small. This simplification is quite
possible.

So, both distributions (in the LS theory and the derived
here) are similar in their dorm. The
similarity is ensured by the narrowness of the tail of
the size spectrum. Namely the situation
of the narrow tail is the dominating one in
the evolution scenario. So, the similarity is the
occasional coincidence corresponding to the
initial
exponential form of the size spectrum tail.
But namely this coincidence
leads to the experimental confirmation of the LS
technique.

Later we return to the situation of the wide tail. Every
wide tail as the narrow tail is also local in the size
axis and, hence, there is the backlash in dependence of
$v_u$ on time. The tail can be approximated in frames
of the steepest descent method by an exponent. So, the
style to construct the solution will remain the same.
Hence, everything is reduced to the already analyzed
situation.

One can see the following stabilizing property -
the wider is the tail, the wider is the
backlash and the local character is approximately
conserved. This property is very important - it is
responsible for the observation of the LS-like
spectrum already at the moderate time.

Here the free molecular regime is adopted, this allows
to write the balance equation in the integral form.
The opposite case is the case of the profiles of
metastability around the embryos. In this case one has
to take into account the interactions between these
profiles. The task seems to be extremely complex.
Nevertheless the answer for the form of the size
distribution is very simple. Certainly this answer is
rather approximate.

Really, one can propose the following model. Since the
profiles are sharp functions of the space coordinate
one can imagine only the pair interactions. Such a
pair battle will end by defeat on one of partners. The
winner will continue to be the embryo of a new phase,
the looser disappears. At the asymptotics of evolution
the remaining embryo had to win many battles. With
probability $p_1$ it wins the first battle, with
probability $p_i$ it wins the $i$-th battle. The total
probability $P_{tot}$ to win all battles is the product
$\Pi_i p_i$ of all probabilities. Since $p_i$ are
independent stochastic values we have for $P_{tot}$
the
log-normal distribution.

Certainly this approach can be spread to the group
interactions (triple, etc.) which will give the same
final result.

\section{Development of the model}

Now we can note several important properties of the
size spectrum. The first property concerns the
influence of the boundary condition on the tail of the
size spectrum. Consider
$$
f=-f_- + f_+
$$
$$
f_- \sim \exp(-(x+x_0)^2/4Dt)
$$
$$
f_+ \sim \exp(-(x-x_0)^2/4Dt)
$$
Then
$$
f_-  \sim \exp(-x^2/4Dt) \exp(-2xx_0/4Dt) \exp(-x_0^2/4Dt)
$$
$$
f_+  \sim \exp(-x^2/4Dt) \exp(2xx_0/4Dt) \exp(-x_0^2/4Dt)
$$
$$
f_+/f_- \sim \exp(4xx_0/4Dt)
$$ and one can take into account in the tail only the
term $f_+$.

One can add that the regular growth can not destroy
the tail - one can speak only about the shift of the
tail and the sequential cut-off of the regions
preceding the tail.

The second property is the possibility to sweep out the
boundaries between stages in the sequential
description of the evolution. Really, does the
diffusion stop after the end of diffusional blurring?
Certainly, it continues and the time of diffusional
blurring depends on the amplitude
of the spectrum, i.e. of the
rest of the tail.

One can note the following important
feature - The
tail blurring is so fast that it can not be overcome
by the regular growth. So, the time of diffusion
blurring is important  and the diffusion process
occurs during  the whole time of evolution.

Now we specify the recipe of calculations for concrete
case. We shall explicitly see what effect has the
relatively small backlash in the law of growth.

\subsection{Explicit calculations}

We start from the law of growth
$$
\frac{d\rho}{dt} = \frac{\zeta}{t_1} (1-u^{-1})
$$
for $u=\rho/\rho_c$. Since
$$
\frac{d\rho}{dt} = \frac{d\rho_c u}{dt} =
\rho_c \frac{du}{dt} + u \frac{d\rho_c}{dt}$$
we see that
$$
\frac{du}{dt} = \frac{1}{\rho_c}
\frac{\zeta}{t_1} (1-u^{-1}) - \frac{u}{\rho_c}
\frac{d\rho_c}{dt}
$$
Since $$ \zeta=\frac{2a}{3\rho_c}$$
we come to
$$
\frac{du}{dt} = \frac{2a}{3\rho_c^2 t_1}
 (1-u^{-1}) - \frac{u}{\rho_c}
\frac{d\rho_c}{dt}
$$
or
$$
\frac{3\rho_c^2 t_1}{2a}\frac{du}{dt}
=
 (1-u^{-1}) -\frac{3\rho_c t_1}{2a}
\frac{d\rho_c}{dt} u
$$
We introduce $\tau$ to have
$$
\frac{2a}{3\rho_c^2(t) t_1} dt = d\tau
$$
and then
$$
\frac{du}{d\tau} =
(1-u^{-1}) - \frac{1}{\rho_c} \frac{d\rho_c}{d \tau} u
$$
or
$$
\frac{du}{d\tau} =
(1-u^{-1}) - \frac{d\ln \rho_c}{d \tau} u
$$
So,
$$
\gamma =\frac{d\ln \rho_c}{d \tau}
$$

Now we find the argument $u_m$ which provides maximum for
the rate of growth $du/d\tau$, i.e. the maximum of the
curve $(1-u^{-1}) - \gamma u$. Having
differentiated $du/d\tau$ on $u$ we have
$$
\frac{d}{du} [(1-u^{-1}) - \gamma u] = u^{-2} - \gamma
$$
Then
$$
u_m = \gamma^{-1/2}
$$
The height of the curve $(1-u^{-1}) - \gamma u$ will
be $$\frac{du}{d\tau}|_{max} = 1-2 u_m^{-1}$$
It has to be zero or some small negative value
$-\delta$. Namely $\delta$ is the backlash.
Then
$$
1-2 u_m^{-1} = - \delta$$
and $$ u_m = \frac{2}{1+\delta}$$
$$\gamma = \frac{(1+\delta)^2}{4}$$

Now we reconstruct the dependence of $\rho_c$ on $t$
based on the known value of $\gamma$. We have
$$
\frac{(1+\delta)^2}{4} = \frac{d\ln \rho_c}{d\tau}$$
or
$$
\frac{3 t_1}{4 a} \frac{d \rho^2}{dt} =
\frac{(1+\delta)^2}{4}
$$
Then
$$
\rho_c^2 \sim \frac{a}{3 t_1} (1+\delta)^2 t
$$
or
$$
(\frac{2a}{3\zeta})^2 \sim \frac{a}{3 t_1} (1+\delta)^2 t
$$
Then the supersaturation satisfies the asymptotic
behavior
$$
\zeta \approx \sqrt{\frac{4 a t_1}{3 (1+\delta)^2 t} }
$$

\subsection{Contradiction in asymptotics}

Now we can see the concrete picture for the
approximate law of growth.

The asymptotics $1-\gamma u$ at big $u$ crosses the
axis $du/d\tau=0$ at
$$u=\gamma^{-1}=\frac{4}{(1+\delta)^2}$$
The level of the backlash $-\delta$ it crosses at
$$u_r=(1+\delta) \gamma^{-1}= \frac{4}{ (1+\delta)} $$

The asymptotics $1-u^{-1}$ for the rate of $u$ growth
at small $u$ crosses the axis at
$$
u=1
$$
and crosses the level of the backlash at
$$
u_l = (1+\delta)^{-1}
$$

So, the approximate rate of growth is constructed. One
has to take into account that the backlash $-\delta$
can be also  the function of time $t$ or $\tau$.

Now we analyze the asymptotic behavior of big $u$
We have
$$
\frac{du}{d\tau} \sim 1-\gamma u \sim - \gamma u
$$
Then
$$
\ln u \sim -\gamma \tau \ \ \ \ \ \ \ \ \ \ \
u \sim \exp(-\gamma \tau)
$$

Now we can explicitly express $\tau$ on $t$ based on
the known asymptotics $\rho_c$. Really,
$$
\frac{2a}{3\rho_c^2 t_1} dt = d\tau
$$
or
$$
\frac{2a}{3(\frac{ a t}{3 t_1}) t_1} dt = d\tau
$$
Then
$$
\frac{2}{t} dt = d\tau
$$
and
$$
  2 \ln t \sim \tau
$$

Then the asymptotics for $u$ in $t-$scale will be
$$
u \sim \exp(-\gamma \tau) \sim t^{-2 \gamma} \sim
t^{-1/2}
$$

Now we can see what will be the diffusional blurring
$\exp(-const \ \ s^2/t) $. We see that
$$
\exp(-const \ \ s^2/t) \sim
\exp(-const \ \ \rho_c^2 u^2/t) \sim
\exp(-const \ \ \frac{a}{3t_1} t t^{-1}/t)
$$
and it seems that the diffusion blurring is the main
effect. It is no more than an error. The reason is
that $s$ does not grow here. Really, $\rho = \rho_c u
\sim t^{1/2} t^{-1/2} = const$ does not grow. This
occurs because we throw away the negligible constant
in the law $du/d\tau = 1-\gamma u$. But the effect of
non-zero growth of $\rho$ manifests in the constant
$1$ in this law. Certainly, it will be lost in
comparison with the leading term. So, we see that the
variables in LS theory are very dangerous. It is
forbidden to choose new irregular variables and then
fulfill the asymptotic analysis.

As for the smallness of diffusional blurring one
can easily see it directly. Since $d\rho/dt =
\zeta/t_1$ and we already know that $\zeta \sim
t^{-1/2}$ then the integration gives $\rho \sim
t^{1/2}$ and $s^2 \sim t^2 \gg t$. This shows the
smallness  of diffusional blurring.

One can take into account the modifications of the model
\begin{itemize}
\item
At the tail of the diffusion gaussian the regular shift
is not very important.
\item
The initial diffusional blurring does not stop at the
beginning of the regular dissolution of the tail but
takes place all time long.
\end{itemize}

In the LS approach it is used that all substance is
in the region less than $u_m$ without justifications. Here
we shall show the analogous fact (all substance is near
$u_m$) explicitly. This fact has to correspond
to the smallness of  the
diffusion blurring. We have to show this smallness.
 Really, the
width $s^2 \sim t$ of the diffusion blurring is many
times less than the critical size $\rho_c \sim
t^{1/2}$ since $s \sim \rho^2$.

\subsection{Modifications of the model}

The balance equation is the main instrument to determine
the evolution of the system. It can be written in the
following form
$$
\rho_c^3 \int_0^{\infty} u^3 \phi(u, \tau) du = const
$$
Precisely speaking one has to add the the
supersaturation as $2a/3\rho_c$ and get
$$
\frac{2a}{3 \rho_c} +
\rho_c^3 \int_0^{\infty} u^3 \phi(u, \tau) du = const
$$
but the first term goes to zero. In any case it is
impossible to determine the critical size from the
last equation having
calculated the integral term in some approximation.

We see that in analysis of the balance equation
lies a dangerous possibility to get wrong
results. This possibility is extremely high in the
LS analysis where the size spectrum has to cancel
divergence of the critical size asymptotic behavior.

To see the behavior of $\rho_c$ we have to establish
the form of $\phi$ at least in the asymptotics. We
write
$$
\phi(u,\tau) \rightarrow \phi_{as} (u,t)
$$
where $\phi_{as}$ is defined at big $u$ from the
gaussian (one can show that the back side gaussian
is not important here).
Namely, we have the following chain of equalities
$$
\phi_{as} (u, \tau) du = f(\rho, \tau) d\rho
$$
$$
f(\rho, \tau) d \rho =  \Psi(s, \tau) ds
$$
$$
\Psi (s, \tau) \sim \exp(-\frac{s^2}{4 D_s t(\tau)})
$$
where $s$ is $\rho^2$, $D_s$ is the diffusion
coefficient over $s$ (known value).

Roughly speaking the problem is solved. But we can not
combine the values of $u$, $\rho_c$ and $\tau$ until
we integrate the law $du/d\tau = (1-u^{-1}) - \gamma
u$ of growth. Fortunately we can not do this with $\gamma$
varying in time. So, it is necessary to introduce
approximations for this law of growth. This has been
done above. Actually we are interested now in the size
spectrum for the values $u>u_r$. Then we have
$$
u=\tilde{u} + \int_{\tau(t_0)}^{\tau(t)}
(\frac{du}{d\tau}) d\tau =
\tilde{u} + \int_{\tau(t_0)}^{\tau(t)}
(1-\gamma u) d\tau
$$
where $\tilde{u}$ is the size of $u$ at $t_0$.
Here $t_0$ is the time when the diffusion transforms
into the regular motion and $\tilde{u}$ is the
corresponding size.
The value of $\tilde{u}$ can be found on the base of
$u,\tau$ and the initial value, then we take the
initial size spectrum at $\tilde{u}$ and get the size
spectrum for $u$ at $\tau$.

Very approximately we can substitute the law
$$
\frac{du}{d\tau} = 1 - \gamma u
$$
by
$$
\frac{du}{d\tau} = - \gamma u
$$
This gives
$$
u = \tilde{u} \exp(-\int_{\tau(t_0)}^{\tau(t)}
\gamma(\tau) d\tau )
$$
or
$$
 \tilde{u} = u  \exp(\int_{\tau(t_0)}^{\tau(t)}
\gamma(\tau) d\tau )
$$
But as we have seen earlier this leads to an error.
Fortunately we can integrate already $du/d\tau=1-\gamma(
\tau) u$ explicitly without simplification
(formulas will be long).

On the base of $\tilde{u}$ one can find
$$
\tilde{\rho} = \tilde{u} \rho_c (t_0)
$$
and then $\tilde s$ equal to $\tilde{\rho}^2$.

For the distribution in $\tilde{\rho}$-scale
 it is very easy to write the
gaussian
$$
G \sim \exp(-\frac{\tilde{\rho}^4}{D_s t(\tau)})
\frac{d\rho}{ds} \approx
\exp(-\frac{\tilde{\rho}^4}{D_s t(\tau)})
$$
Here we ignore the jacobians arrived from transition
from $s$ to $\rho$ scale because $\exp(-const x^4)$ at
the tail
is a very sharp function. We shall ignore them below
also.

We are interested now in behavior at $u \approx u_r$.
Here the true approximation will be
$$
\frac{du}{dt} = c_1 - \gamma u
$$
$$
c_1 = 1- u_r^{-1}
$$
Let us take $\gamma \approx \gamma_B =
\gamma(t_B)$ ($t_B$ is the
time when $u_r$ is attained).

Then
$$
\ln[\frac{u-c_1/\gamma_B}{\tilde{u}-c_1/\gamma_B}] =
-[\tau(t) -\tau]\gamma_B
$$
So,
$$
u-c_1/\gamma_B =(\tilde{u}-c_1/\gamma_B)
\exp(-\gamma_B (\tau(t) - \tau(t_0)))
$$
and $\tilde{u}$ is $u$ at $t_0$. As the result we know
$\tilde{u}$ as a function of $u$, i.e. $\tilde{u} = F(u)$.

In any case we can integrate the equation with $\gamma
= \gamma(\tau
)$ since we have the first order linear differential
equation. This will give the real true result with
rather long formulas. So, we use the previous formulas
keeping in mind the necessity to apply the formulas
with varying $\gamma$.

Now we can see the asymptotics for the distribution
function
$$
\phi(u,\tau) \sim
\exp( - \frac{\tilde{u}(u)^4 \rho_c^4}{D_s t})
\frac{c_1 - \gamma_B \tilde{u}(u)}{c_1 - \gamma_B u}
$$
This result is for $u \geq u_r$.

One of important properties of this solution is the
weak dependence of the form (after scaling)
of the size spectrum on $\tau$. The quantity of
substance in the
region $[u_r, \infty)$ can be found as
$$
G_> = \rho_c^3 \int_{u_r}^{\infty} u^3 \phi(u,\tau) du
$$

Now we shall analyze the region $[u_l, u_r]$. The
values at this region are marked by the subscript $=$.
Here
$$
\frac{du}{d\tau} = - \delta
$$
Then
$$
\phi(u,\tau) = \phi_=(u, \tau) =
\phi(u_r,
\hat{\tau})\frac{\frac{du_r(\hat{\tau})}{d\tau}}
{\frac{du(\tau)}{d\tau}}
$$
where $\hat{\tau}$ is defined as
$$
u = \hat{u} - \int_{\hat{\tau}}^{\tau} \delta (\tau') d\tau'
$$
which is a rather complex closure. The previous equation can
be rewritten as
$$
\phi(u,\tau) = \phi_=(u, \tau) =
\phi(u_r,
\hat{\tau})\frac{\delta(\hat{\tau})}
{\delta(\tau)}
$$

More rigorous is to make a shift of $u_r$ to exclude a
slow evolution right of $u_r$. But this does not
change the qualitative behavior.

The quantity of substance $G_=$ in this region is
given by
$$
G_=  = \rho_c^3
\int_{u_l}^{u_r} u^3 \phi_=(u,\tau) du
=\rho_c^3
\int_{u_l}^{u_r} u^3
\phi(u_r,
{\tau_r})\frac{\delta({\tau_r})}
{\delta(\tau)} du
$$

Here
$$
u_r = u + \int_{\tau_r}^{\tau} \delta (\tau") d \tau"
$$

Now we investigate the region $[0,u_l]$. The values at
this region will be marked by the subscript $<$. Here
the size spectrum can be given by
$$
\phi(u,\tau) = \phi(u_l, \tau')
\frac{\delta(\tau')}{1-u^{-1}}
$$
The quantity of substance is given by
$$
G_< = \rho_c^3 \int_0^{u_l} u^3 \phi(u_l, \tau')
\frac{\delta(\tau')}{1-u^{-1}} du
$$
Here
$$
\tau - \tau' = \int_u^{u_l} \frac{du}{1-u^{-1}}
$$

The unknown function is $\delta(t)$.

All quantities of substance $G_<$, $G_=$, $G_>$  have
to be substituted into the balance equation which
gives the  equation on $\delta(t)$ with the known coefficients.

\subsection{Steepest descent procedure}

Now we shall discuss the ways to solve this equation.
At first we have to get a true algebraic equation. We
differentiate the balance equation on time and get
$$
\frac{d(G_<+G_=+G_>)}{d\tau} =0
$$
Then we use concrete approximations
to get $dG_>/d\tau$,
$d G_</ d\tau$,
$dG_= / d\tau$.

For $dG_>/d\tau$ we see that
 the subintegral function is the
product of the three functions:
\begin{enumerate}
\item
 the moderate
function $3 u^2$,
\item
 the rapidly growing function for
the absolute value of the rate of growth $du/d\tau$.
This function becomes very small at $u_r$ (practically
it is zero).
\item
 the rapidly decreasing function for the initial size
spectrum. This function decreases in the main term
even faster than
$\exp(-x^4)$ and even being
multiplied by the accelerating rate
of growth and by $u^2$ the product goes to zero at big $u$.
\end{enumerate}

So, one can effectively use the steepest
descent method.

For $dG_=/d\tau$ we have a simple expression which
is approximately proportional to the size spectrum
multiplied on $3 u^2 \delta$ and due to the rapid
decrease of the size spectrum we see that the subintegral
function is the rapidly decreasing function of $u$.
This is the ideal
situation for the application of the steepest descent
method with the maximum at the boundary point (here it
is $u_l$).

For $dG_< / d\tau$ we have the integral with the
subintegral function at the interval $]0,u_l]$,
which is the product of three
rapidly varying functions:
\begin{enumerate}
\item The function
$3 u^2$ which goes to zero at $u=0$,
\item
The function $du/d\tau
\sim 1-u^{-1}$ which goes to infinity at $u=0$ while
$u^2 (du/d\tau)$ goes to zero at $u=0$,
\item the rapidly decreasing size spectrum.
\end{enumerate}

This provides good conditions for the application
of the mentioned steepest descent method with the
point of decomposition inside the interval.

The problem which can appear here is a too sharp form of
the size spectrum which can cause the maximum of the
subintegral function at $\rho=u\rho_c$ which is too
small. Fortunately when such sharp size spectrum will
come to the region $[0,u_l]$ the decrease of the
substance in embryos will cause the decrease of
$\rho_c$. It means that until the diffusion blurring
the spectrum will not be dissolved. This situation is
typical for the formation of the new head which has
been discussed in \cite{pred}.

The mentioned approximations lead to the algebraic
equation on $\delta$. This equation allows the further
simplification if we decompose $\delta(\tau)$ in
Taylor's series on inverse time $\xi=\tau^{-1}$
$$
\gamma(\xi) = \gamma_0+ \sum_j a_j\xi^j
$$
with coefficients $a_j$.
The choice of $\xi$ as a variable is recommended by
the structure of the correction term in the LS
procedure.
Also one can use decomposition near some value $\xi_0$
$$
\gamma(\xi) = \gamma(\xi_0)+ \sum_j a_j(\xi-\xi_0)^j
$$
The last modifications make the analysis of the
evolution a technical task.

\section{Situation of wide tails}

Now we shall analyze the situation of the wide tails of the size
spectrum. It means that the half-width of the size
spectrum in the region $\rho>\rho_r$ is many times greater
than $\rho_c$. It is clear that the approximate law of
growth $du/d\tau=1-\gamma u$ corresponds to the law of
growth $d\rho/dt = \zeta/t_1$ and the distribution
function moves as a whole along $\rho$-axis. The
critical size $\rho$ moves faster (but with the same
time dependence $t^{1/2}$) and eats the spectrum
sequentially. For all distribution tails (here we take the
tail multiplied by the number of molecules in the droplet
$\rho^3$) which
decrease like $1/\rho^i$ with $i$ greater than $1$
the relative half-width
will decrease. But the situation with $i<1$ do not
ensure the finite substance, it is forbidden situation.
So, earlier or later the tail will be narrow.
One can speak, thus, only about the intermediate
asymptotics.

Since here the width remains greater than the critical
radius one can speak about another asymptotic behavior,
at least the intermediate asymptotic corresponding to
the wide spectrum can be observed.

Kinetics of the process here will differ from the LS
case. All the time the main consumers of the metastable
phase substance will be the big embryos and the
backlash here is very wide. The asymptotic
$$ \tau(u) = \gamma \ln u
$$
ensures the infinite time of the embryos dissolution
for the infinitely big embryos. Then the main
supposition in the LS theory fails. This argument
states that to
keep the balance of the substance it is impossible that
some relative size will only grow and it is impossible
that all relative sizes will decrease.
But now it is possible to see here the situation where
all sizes decrease rather intensively but the main
consumers of the vapor are the big embryos with
$u$ many times greater than $u_m$.
In the zero approximation the evolution is
very simple - the size spectrum is cut-off by the
critical size (for the sizes less than the critical one
the dissolution is so rapid that we can speak about the
instantaneous dissolution and neglect this region).
This cut off corresponds to the conservation of the
substance in the system.

Define that here  the size spectrum $f_{long}(\rho)$
has the characteristic width
$\Delta (f_{long})$ determined in the integral way as
$$
\int_0^\infty f_{long} (\rho) d\rho =
f_{long}^{max}\Delta (f_{long})
$$
where $f_{long}^{max}$ is the amplitude of the size
spectrum, or in the differential way as
$$
f_{long}^{max}(\rho_m + \Delta (f_{long})) =
f_{long}^{max}/\exp(1)
$$
where $\rho_m$ is the argument for the amplitude value
of spectrum.
This width satisfies the strong inequality
$$
\Delta (f_{long}) \gg \rho_c
$$
The substance balance here is written as
$$
\Phi = \zeta +q_+
$$
where $\Phi$ is the supersaturation without the
embryos formation, $q_+$ is the substance in the tail
which can be calculated as
$$
q_+ = \int_{\rho_c}^\infty f_{long}(\rho) \rho^3 d\rho
\approx
\int_{(2\div3)\rho_c}^\infty f_{long}(\rho) \rho^3 d\rho
$$
Then in the last integral we can take for
$f_{long}$ the size spectrum fully determined by the
supercritical law of growth from the "initial"
spectrum $f_0$ (at the time $t_{init}$):
$$
f_{long}(\rho, t) = (\zeta(t_{init})/\zeta(t))
f_0(\hat{\rho}(t,t_{init}))
$$
where
$\hat{\rho}(t,t_{init})$ is determined from
$$
\rho= \int_{t_{init}}^t \zeta(t') / t_1 dt' +
\hat{\rho}(t,t_{init})
$$

It is more convenient to study the initial spectrum
and to see how much this spectrum is cut off. So, we
introduce the initial $\rho$-size variable $s$ and write a
balance equation
$$
\Phi = \zeta(t) + \int_{s_c}^\infty \varphi(s)^3 f_0(s)
ds
$$
Here $s_c$ is the initial size of the variable
$s$ which attains at
$t$ the size $(2\div 3)\rho_c$, $\varphi(s)$ is the
size which will be attained at $t$ by the embryo with
initial size $s$
$$
\varphi (s) = s + \int_{t_{init}}^t
\frac{\zeta(t')}{t_1} dt'
$$

The balance equation can be rewritten as
$$
\Phi = \zeta(t) + \int_{s+\int_{t_{init}}^t
\frac{\zeta(t')}{t_1} dt' > (2 \div 3)
\frac{2a}{3\zeta(t)}
} \varphi(s)^3 f(s)
ds
$$
or since $(2 \div 3)
\frac{2a}{3\zeta(t)}$ has to be many times smaller
than the width of the spectrum then
$$
\Phi = \zeta(t) + \int_{s+\int_{t_{init}}^t
\frac{\zeta(t')}{t_1} dt' > 0
} \varphi(s)^3 f_0(s)
ds
$$
Having noticed that $s$ has to be at least positive we
get the following modification of the balance equation
$$
\Phi = \zeta(t) + \int_{s > 0
} \varphi(s)^3 f_0(s)
ds
$$
This equation can be easily solved. Having introduced
the explicit equation for $\varphi$ we come to
$$
\Phi = \zeta(t) + \int_{s > 0
} (s + \int_{t_{init}}^t
\frac{\zeta(t')}{t_1} dt')^3 f_0(s)
ds
$$
One can see that the integral term is the
polynomial
on
$$
\rho_m = \int_{t_{init}}^t
\frac{\zeta(t')}{t_1} dt'
$$
Then
$$
\Phi = \zeta(t) + \sum_{i=0}^3 \frac{3!}{i! (3-i)!}
a_i \rho_m^{3-i}
$$
with
known constants
$$
a_i = \int_0^\infty s^i f_0(s) ds
$$
It can be rewritten as
$$
\Phi = t_1 \frac{d\rho_m}{dt}
+ \sum_{i=0}^3 \frac{3!}{i! (3-i)!}
a_i \rho_m^{3-i}
$$

The last equation is the  differential Abel equation -
the ordinary first oder differential equation. Since
there is no explicit dependence on the argument this
equation can easily integrated. This gives the
solution of the problem.

Certainly, the last solution is not accurate because
there all embryos remain in the integral term - they
remain supercritical ones. This leads to the
qualitatively wrong behavior. It is clear that the
error will be essential namely when the tail stops to
be really the wide tail. But nevertheless it is
possible to take into account the dissolution of the
embryos. The balance equation has to be written as
$$
\Phi = \zeta(t) + \int_{s > (2\div 3) 2a/3\zeta
} (s + \int_{t_{init}}^t
\frac{\zeta(t')}{t_1} dt')^3 f_0(s)
ds
$$
and conserves the polynomial structure on $\rho_m$.
$$
\Phi = \zeta(t) + \int_{s > (2\div 3) 2a/3\zeta
} (s + \rho_m)^3 f_0(s)
ds
$$
or
$$
\Phi = \zeta(t) + \sum_j \rho_m^j
\frac{3!}{j!(3-j)!} \int_{s > (2\div 3) 2a/3\zeta
} s^{3-j}  f_0(s)
ds
$$
Now the coefficients
$$
a_j =\frac{3!}{j!(3-j)!} \int_{s > (2\div 3) 2a/3\zeta
} s^{3-j}  f_0(s)
ds
$$
are known (since the initial size spectrum is known)
functions of $\zeta$. But it is possible to solve this
equation on $\rho_m$ as the third power algebraic
equation
$$
\rho_m = F(\zeta)
$$
or
$$
\rho_m = F(t_1 \frac{d\rho_m}{dt})
$$
with a known function $F$
With the help of the inverse function $F^{-1}$ we can write
$$
\frac{d\rho_m}{dt} = t_1^{-1} F^{-1}(\rho_m)
$$
The last equation can easily solved.

Now we take into account the growth of the embryos.
The balance equation has to be written as
$$
\Phi = \zeta(t) + \int_{s +\rho_m > (2\div 3) 2a/3\zeta
} (s + \int_{t_{init}}^t
\frac{\zeta(t')}{t_1} dt')^3 f_0(s)
ds
$$
and does not conserve the polynomial structure on $\rho_m$.
It can be written analogously
$$
\Phi = \zeta(t) + \sum_j \rho_m^j
\frac{3!}{j!(3-j)!} \int_{s +\rho_m > (2\div 3) 2a/3\zeta
} s^{3-j}  f_0(s)
ds
$$ or
in
the standard form
but with the coefficients
$$
a_j =\frac{3!}{j!(3-j)!} \int_{s+\rho_m > (2\div 3) 2a/3\zeta
} s^{3-j}  f_0(s)
ds
$$
depended (since the initial size spectrum is known)
on $\rho_m$ and $\zeta=t_1 d\rho_m/dt$.
So, we have the first order differential equation
without explicit dependence on the argument $t$
$$
\Psi(\rho_m, d\rho_m/dt) =0
$$
This equation can be integrated when we can express
$d\rho_m/dt$ via $\rho_m$. This is an algebraic
problem which can be solved at least locally in a good
approximation.

\section{Approximations in the explicit
construction the size spectrum}

The first task in construction of the size spectrum is
the construction of initial distribution which will be
later gradually dissolved during the
over-condensation. This task is very complex. Even the
elementary approximate blocks for solution do not allow the
solution. We do
not know the solution of the diffusion equation
for the diffusion blurring of the spectrum even
with the stationary value of the supersaturation.
As the result of such difficulties only very
approximate methods can be formulated.

To see the initial distribution it is more easy to use
the $\rho$-scale because in this scale the asymptotic
law of growth is rather simple and the velocity of
growth contrary to the $u$-scale does not go to
infinity.

The first approximate model, which allows solution is
"the model of sequential evolution". We consider the
period of
the initial diffusion blurring.
Here we have the stationary supersaturation (and
the critical radius).

Consider the value of distribution at some $\rho_f$,
time being fixed. Let the initial distribution be
$\delta$-function at $\rho=\rho_c$.
The route from $\rho_c$ to $\rho_f$ will be split
between the diffusion blurring and the regular growth.
In this model the blurring occurs up to $\rho$ equal
to some parameter $\rho_b$. Later there will be the
pure regular growth. Parameter $\rho_b$ is reasonable
to put equal to $2 \rho_c$.

The time of the regular growth up to $\rho_f$ will be
$(\rho_f-\rho_b) t_1 / \zeta$ or
$(\rho_f-\rho_b) t_1 3 \rho_c / 2 a$. Then the time to
quit  the diffusion blurring will be
$$
t_q = t-(\rho_f-\rho_b) t_1 3 \rho_c / 2 a
$$
The supercritical regular growth is the simple
translation of the size spectrum, then we have to
calculate the spectrum at $\rho_b$ and $t_q$ after the
pure diffusion blurring (here for simplicity we do not
consider the back side input $f_-$)
$$
\vartheta(\rho_f,t) \sim
\frac{2\rho_b}{\sqrt{4D_st}}
\exp(-\frac{\rho_b^4}{4D_st_q})
=
\frac{2\rho_b}{\sqrt{4D_st}}
\exp(-\frac{\rho_b^4}{4D_s(
t-(\rho_f-\rho_b) t_1 3 \rho_c / 2 a)})
$$

The function $\vartheta$ can be considered as the
initial size spectrum. The problem is, thus, solved.

But the function $\vartheta$ has a certain
disadvantage - the size spectrum is finite. Really,
for $\rho_f$ greater than $\rho_{lim}$
$$
\rho_{lim} = \rho_b + 2 a t/ t_1 3 \rho_c
$$
the spectrum is zero.

Actually the values of $\rho$ near $\rho_{lim}$ begin
to dissolve (i.e. $\rho_{lim}$ is near $\rho_r$) when
the amplitude of the rest of the spectrum is extremely
small. So, the relative quantity of droplets in
negligibly small. It will be big only in the systems
of cosmological sizes.

Nevertheless one can refine this solution. Fortunately,
the diffusion process can be easily estimated even
with the varying $\rho_c$ because here $\zeta$ is very
small and the average number of collisions in the time
unit will be $(2/t_1)(3\rho^2(t))$. Here we can take
$\rho$ on the base of the regular growth and get the
total number of collisions
$$
n_{tot} = \int_{0}^{t} 6 \rho^2(t') dt' /t_1
$$
The characteristic width $\Delta_{tot}$
can be estimated as
$\Delta_{tot} =\sqrt{2 n_{tot}}$ with a
sufficient accuracy.

Then the resulting distribution will be proportional
to
$$
\hat{\vartheta} =
\int_{-\infty}^{\infty}
\exp(-\frac{(\rho-\rho_f)^2}{2 n_{tot}})
\vartheta(\rho_f,t) d\rho_f
$$
or
$$
\hat{\vartheta} =\frac{2\rho_b}{\sqrt{4D_st}}
\int_{-\infty}^{\infty}
\exp(-\frac{(\rho-\rho_f)^2}{2 n_{tot}})
\exp(-\frac{\rho_b^4}{4D_s(
t-(\rho_f-\rho_b) t_1 3 \rho_c / 2 a)})
 d\rho_f
$$
It is quite satisfactory here
to consider only the right hand wing
of the distribution, i.e. to put $\rho>\rho_f$. Since
the subintegral function is the product of exponents
it is reasonable to use the steepest descent method.

The required result is attained by a simple
combination of the regular growth and the pure
diffusion. The cross effects when, for example,
the stochastic
increase of the embryos size leads to increase of the
regular rate of growth are missed here. They can be
included into consideration by consideration of the
effective half-width of the diffusion blurring and the
shift in regular growth proposed in \cite{stohgr}.
Certainly, they have to be slightly reconsidered since
now we start not from the very beginning of the size
axis.

Now we shall show the primitive approximate way to
construct the explicit form of the size spectrum for
the narrow tail.

The initial distribution is supposed to be known.

We approximate the rate of
growth for $u$ in rescaled time $\tau$
by the
following approximation
$$
\frac{du}{d\tau} \approx v_m \equiv max \{ \frac{du}{d\tau}
\} = - \delta(\tau)
$$
for $u \in [u_l, u_r]$
$$
\frac{du}{d\tau} \approx 1- \gamma u
$$
for $u>u_r$,
$$
\frac{du}{d\tau} \approx 1-u^{-1}
$$
for $u<u_l$.

Parameters $u_l$ and $u_r$ have to be chosen to ensure
the continuity of the whole approximation.

Consider $u>u_r$. The law of growth for $\rho$
corresponding to $du/d\tau \sim (1-\gamma u)$ is
$d \rho / dt = \zeta/t_1 $ (the r.h.s. is precisely taken
into account  by transition from $t$ to $\tau$). So,
the distribution in $\rho$-scale is moving as a whole
without changing of the form. We know this form - it
is exponential form
$$
f(\rho) \sim \exp(-\alpha \rho)
$$
with some parameter $\alpha$.

Then the distribution $\varphi$ over $u$ is connected
with $f$ by the following relation
$$
\varphi (u) = f(\rho)
\frac{\frac{d\rho}{dt}}{\frac{du}{dt}}
$$
Having inserted the explicit relations for the
derivatives we come to
$$
\varphi (u) = f(\rho)
\frac{1-u^{-1}}{(1-u^{-1})-\gamma u}
$$
or in the supercritical limit
$$
\varphi (u) = f(\rho)
\frac{1}{1-\gamma u}
$$

Here the distribution function is even sharper than
the distribution over $\rho$.

Now at first
we shall assume that the backlash is changing in
time slowly
in comparison with the time of dissolution
of an embryo from $u_r$ to $u_l$.
This case will be at least the base approximation for
iteration procedures to refine the solution.

The known solution in the region $u>u_r$ leads to the
known rate of appearance $\Psi_b(\tau)$ at $u=u_r$. This
value is given by
$$
\Psi_b (\tau) = \varphi (u)|_{u=u_r}
(\frac{du}{d\tau}|_{u=u_r}) =
\varphi (u)|_{u=u_r} v_m
$$
Here $\varphi$ is the distribution over $u$. Now it
can be established. One can easily show that
$\Psi_b(\tau)$ at $u=u_r$ is sharp decreasing function
of $\tau$.

The last function is the source at the left side of
the central interval $[u_l, u_r]$. Now it is possible
to solution at this interval. This solution is very
simple and it is given by
$$
\varphi(u,\tau) =
\Psi_b(\tilde{\tau})/\frac{du}{d\tau}|_{u=u_r}
$$
or
$$
\varphi(u,\tau) =
\Psi_b(\tilde{\tau})/v_m
$$

The time $\tilde{\tau}$ satisfies relation
$$
u-u_r = \int_{\tilde{\tau}}^{\tau} v_m(t') dt'
\approx v_m (\tau-\tilde{\tau})
$$
Hence, the solution in the central region is
constructed.

Generalization for the varying $\delta$ is rather
simple.
The  rate of appearance $\Psi_b(\tilde{\tau})$
at $u=u_r$ is given by
$$
\Psi_b (\tilde{\tau}) =
\varphi (u|\tilde{\tau})|_{u=u_r} v_m(\tilde{\tau})
$$
Here $\varphi$ is the distribution over $u$. Now it
can be established.

The last function is the source at the left side of
the central interval $[u_l, u_r]$. Now it is possible
to solution at this interval. This solution is very
simple and it is given by
$$
\varphi(u,\tau) =
\Psi_b(\tilde{\tau})/v_m(\tau)
$$
and
$$
\tilde{\varphi}(u,\tau) =
\varphi (u| \tilde{\tau})|_{u=u_r(\tilde{\tau})}
v_m(\tilde{\tau})/v_m(\tau)
$$

The time $\tilde{\tau}$ satisfies relation
$$
u-u_r = \int_{\tilde{\tau}}^\tau v_m(\tau') d\tau'
\neq v_m (\tau-\tilde{\tau})
$$

Now only the last region - the region of small
$u<u_l$, has to be investigated. Here the solution is
also rather simple - it is the simple translation of
the source from $u_r$ under the law of
growth independent on $\gamma$. Here all constructions are
analogous to the previous case but the transition over
the central region has to be taken into account.

The distribution $\varphi$ is given by
$$
\varphi(u,t) \sim
\Psi_a(\hat{\tau})/(du/d\tau)
\sim \Psi_a(\hat{\tau})/(1-u^{-1})
$$
where
$$
\Psi_a(\hat{\tau}) = \tilde{\varphi}(u_l(\hat{\tau}),
\hat{\tau})  v_m(\hat{\tau})
$$
Here $\hat{t}$ is determined by the following
way
$$
\hat{\tau}-\tilde{\tau} = \int_{u_l}^{u_r}
\frac{1}{v_m} du
$$
and
$$
\tau-\hat{\tau} = \int_{u}^{u_l}
\frac{1}{du/d\tau} du
$$

For $\delta = const $ one can simplify the last relation
$$
\tau-\tilde{\tau} = \int_u^{u_l} \frac{1}{du/d\tau} du +
(u_l-u_r)/v_m
$$
or
$$
\tau-\tilde{\tau} = \int_u^{u_l} \frac{1}{1-u^{-1}} du +
(u_l-u_r)/v_m
$$
The integral can be easily taken analytically and we get
the explicit expression for $\tilde{t}$.

If $v_m$ does not essentially depend on time one can
simply solve all these equations.
The form of the size spectrum is determined, only
parameter $v_m$ ($\gamma$, $u_r$, $u_l$ depend on
$v_m$) is unknown. The balance equation will be
algebraic equation on $v_m$ and can be easily solved
since we know at least the zero approximation (for example,
one can take LS asymptotic).

It is clear that the backlash is not a constant value,
it changes in time. The effective way to investigate
the situation of the varying backlash is to consider
this variation small, then to decompose $v_m(\tau)$ in
Taylor's series, to take few first derivatives and to
fulfill the same program as was done above in the case
of the constant value $v_m$.

\section{Finite number of embryos}

Under the finite spectrum of sizes the LS asymptotic
will be also destroyed.
After the size of the greatest embryo attains the
critical value and $\gamma$ goes to zero
the diffusion blurring begins and we return to
the section about the diffusion blurring and the
evolution will be described by the same formulas
as mentioned above.
Here it is important that the diffusion term plays the
main role in evolution. Certainly, diffusion begins to
play essential role
earlier than the biggest embryo attains the
critical size.

We see that the further scenario is built on the
doubtful alternative – whether the size spectrum finite
or not. The diffusion process through the formula for
the Green function gives the infinite size spectrum.
But is every concrete system the spectrum is the finite
one. The answer on this question also determines the
asymptotic. The type of asymptotic is determined by the
time of observation and the sizes of the system under
the observation. Fortunately this question is
artificial because other effects (the change of the
regime of growth, the thermal effects, etc.) lead to
the end of applicability of the chosen physical model.

In every system the finite spectrum is the direct
consequence of the finite number of embryos. So, we
have to develop methods to describe the evolution with
the finite number of embryos.

Suppose we have few embryos in the system. Then the
evolution is determined by the laws of their regular
growth (diffusion has also to be taken into account but
in the manner of some stochastic adsorption and
ejection of the molecules by the embryo). The balance
equation links these laws of growth in the closed
system of equations. The number of these equations
equals the number of embryos. When the number of
embryos is less or equal to few hundreds it is
preferable to solve these equations by computers
explicitly.

We rewrite the law of growth for a chosen embryo in the
following form
$$
\frac{d\rho_i}{dt} = \frac{\zeta}{t_1}(1-\frac{\rho_c}{\rho_i})
$$
$$
\rho_c \equiv \frac{2a}{3\ln(\zeta+1)}
$$
(index $i$ marks embryos), where $a$ is the
rescaled surface tension (it is a constant) and
$\zeta$ is the supersaturation. Namely through the
supersaturation one can link the laws of growth having
written the balance equation
$$
\zeta = \Phi - \sum_i \rho_i^3
$$
Here $\Phi$ is the initial value of supersaturation.

One can easily solve the system of these equations at
least approximately. The first method is very simple.
At given initial sizes of embryos we find the value of
the supersaturation. Then we reconsider the sizes
having moved them according to the law of growth at the
given supersaturation. The time interval has to be
small. Then we recalculate the supersaturation having
made one step of evolution. This method is the step by step
procedure.

The stochastic adsorption and ejection of the
molecules by the embryo can be taken into account by
the following simple procedure. We keep in the memory
of computer all coordinates of embryos $\rho_i$
and for every
embryo at the given supersaturation we determine the
rate of adsorption as
$$
R_+ = 3 \rho^2 (\zeta+1) \Delta t / t_1
$$
Here $\Delta t$ is an elementary time step.
The rate of ejection will be
$$
R_- = R_+ - (d\rho/dt) \Delta t
$$
$$
\frac{d\rho}{dt} =
\frac{\zeta}{t_1}(1-\frac{2a}{3\zeta\rho})
$$
The rate of staying still will be
$$
R_0 = 1-R_- - R_+
$$
We must choose $\Delta t$ enough small to have $R_0>0$
even for the greatest embryo.

Having put the point at the interval $[0,1]$
stochastically we determine what action we have to do.
If this point belong to interval $[0,R_-]$ we eject
the molecule, i.e. we make a transition $\rho_i
\rightarrow \rho_i-1$. If this point belong to
interval $[R_-,R_- + R_+ ]$ we accumulate
the molecule, i.e. we make a transition $\rho_i
\rightarrow \rho_i+1$. If this point belong to interval
$[R_-+R_+,1]$ we keep the coordinate still,
i.e. we make a transition $\rho_i
\rightarrow \rho_i$.

We have to repeat this action for every embryo. Then
we recalculate the supersaturation
$$
\zeta = \Phi - \sum_i \rho_i
$$
and fulfill the step in time.

We see that these procedures can not give the
analytical properties of evolution with the finite
number of embryos.
Hence, the problem of the system of
several embryos exists and the effective solution is absent.

Now we turn to the simplest case - the case of small
quantity of embryos.
The asymptotic of the process is the evident – there
will
be the greatest embryo which will be the critical one.
This embryo is in the effective
potential well. Description here
is analogous to the case of the several identical
embryos.

Really, the law of growth
$$
\frac{d\rho}{dt} =
\frac{\Phi- \rho^3}{t_1}(1-\frac{2a}{3 \ln (\Phi - \rho^3
+1)\rho})
$$
corresponds to the regular motion in the potential
$$
U =\int_0^\rho \frac{\Phi - \rho^3}{t_1}(1-\frac{2a}{3
\ln (\Phi -
\rho^3 +1)\rho}) d\rho
$$
 which is very deep and only in the region of small $\rho$
it has a barrier and begins to decrease when $\rho$
decreases.

If we ignore the fluctuational formation of the new
embryos then the evolution leads to the fluctuational
disappearance of the last embryo. It requires
absolutely giant times and at these times new
supercritical embryos have to appear. So, the ignorance
of the fluctuational formation of
new embryos (with extremely slow rate) is illegal.

The flow of disappearance of the last embryo is many
times less than the flow of appearance (formation) of
the new embryo even in the system with the practically
consumed metastable phase by the giant embryo. So, one
can see the formation of the second embryo. Later one
can observe the competition between these embryos.
Practically inevitably the second (new) embryo will be
dissolved but with a small probability $\pi$ which can
be estimated (very roughly because we keep the old
potential $U$ appropriate only for one embryo) as
$$
\pi \sim \exp (U(\rho=\rho_{min}/2^{1/3}) - U(\rho_{min}))
$$
(here $\rho_{min}$ is the argument of the minimum of
potential) the first embryo becomes the second one and
it will be dissolved. Certainly, the above presented
picture is only the rough estimate.

\section{Conclusions}

In the theoretical constructions presented above we
came to the following results
\begin{itemize}
\item
In Sections 2,3 we pointed out the weak features of
the LS approach. In section 2 the features concerning
the behavior of the supersaturation  were outlined. In
section 3 the weak features in construction of the
size spectrum were presented.
\item
In section 4 the sequential analysis of the evolution
was given and it is shown why the form of the size
spectrum resembles the form given in the LS approach
\item
In section 5 the details of the most difficult and the
most important period
of the tail dissolution are given
\item
In section 6 the approximate way to construct the
initial size spectrum for the period of the tail
dissolution is given. This point is very important for
qualitative results.
\item
In section 7 the situation of the wide tail which
allows essential simplification is presented
\item
Section 8 is devoted to the specifics of the case when
only few embryos remain in the system.
\end{itemize}

Sections 7 and 8 are supplementary ones, the complete
theory is given in sections 4 - 6. This theory gives
an answer on two important questions - why LS spectrum
of sizes can be really seen in nature and what is the
difference between the real situation and the LS
approach.

\end{document}